\documentstyle[12pt]{article}

\catcode`\@=11
\long\def\@makefntext#1{
\protect\noindent \hbox to 3.2pt {\hskip-.9pt
$^{{\ninerm\@thefnmark}}$\hfil}#1\hfill}     %CAN BE USED

\def\@makefnmark{\hbox to 0pt{$^{\@thefnmark}$\hss}}  %ORIGINAL

\def\ps@myheadings{\let\@mkboth\@gobbletwo
\def\@oddhead{\hbox{}
\rightmark\hfil\ninerm\thepage}
\def\@oddfoot{}\def\@evenhead{\ninerm\thepage\hfil
\leftmark\hbox{}}\def\@evenfoot{}
\def\sectionmark##1{}\def\subsectionmark##1{}}

\renewcommand{\thefootnote}{\fnsymbol{footnote}}

\newcounter{sectionc}\newcounter{subsectionc}\newcounter{subsubsectionc}
\renewcommand{\section}[1] {\vspace*{0.5cm}\addtocounter{sectionc}{1}
\setcounter{subsectionc}{0}\setcounter{subsubsectionc}{0}\noindent
   {\normalsize\bf\thesectionc. #1}\par\vspace*{0.4cm}}
\renewcommand{\subsection}[1] {\vspace*{0.6cm}\addtocounter{subsectionc}{1}
   \setcounter{subsubsectionc}{0}\noindent
   {\normalsize\it\thesectionc.\thesubsectionc. #1}\par\vspace*{0.4cm}}
\renewcommand{\subsubsection}[1]
{\vspace*{0.6cm}\addtocounter{subsubsectionc}{1}
   \noindent {\normalsize\rm\thesectionc.\thesubsectionc.\thesubsubsectionc.
   #1}\par\vspace*{0.4cm}}

\newcounter{appendixc}
\newcounter{subappendixc}[appendixc]
\newcounter{subsubappendixc}[subappendixc]

\renewcommand{\appendix}[1] {\vspace*{0.6cm}
        \refstepcounter{appendixc}
        \setcounter{figure}{0}
        \setcounter{table}{0}
        \setcounter{equation}{0}
        \renewcommand{\thefigure}{\Alph{appendixc}.\arabic{figure}}
        \renewcommand{\thetable}{\Alph{appendixc}.\arabic{table}}
        \renewcommand{\theappendixc}{\Alph{appendixc}}
        \renewcommand{\theequation}{\Alph{appendixc}.\arabic{equation}}
        \noindent{\bf Appendix \theappendixc #1}\par\vspace*{0.4cm}}

\renewenvironment{thebibliography}[1]
   {\begin{list}{\arabic{enumi}.}
   {\usecounter{enumi}\setlength{\parsep}{0pt}
\setlength{\leftmargin 1.25cm}{\rightmargin 0pt}
    \setlength{\itemsep}{0pt} \settowidth
   {\labelwidth}{#1.}\sloppy}}{\end{list}}

\newcounter{itemlistc}
\newcounter{romanlistc}
\newcounter{alphlistc}
\newcounter{arabiclistc}

\newcommand{\fcaption}[1]{
        \refstepcounter{figure}
        \setbox\@tempboxa = \hbox{\footnotesize Fig.~\thefigure. #1}
        \ifdim \wd\@tempboxa > 6in
           {\begin{center}
        \parbox{6in}{\footnotesize\baselineskip=12pt Fig.~\thefigure. #1}
            \end{center}}
        \else
             {\begin{center}
             {\footnotesize Fig.~\thefigure. #1}
              \end{center}}
        \fi}

\newcommand{\tcaption}[1]{
        \refstepcounter{table}
        \setbox\@tempboxa = \hbox{\footnotesize Table~\thetable. #1}
        \ifdim \wd\@tempboxa > 6in
           {\begin{center}
        \parbox{6in}{\footnotesize\baselineskip=12pt Table~\thetable. #1}
            \end{center}}
        \else
             {\begin{center}
             {\footnotesize Table~\thetable. #1}
              \end{center}}
        \fi}

\def\@citex[#1]#2{\if@filesw\immediate\write\@auxout
   {\string\citation{#2}}\fi
\def\@citea{}\@cite{\@for\@citeb:=#2\do
   {\@citea\def\@citea{,}\@ifundefined
   {b@\@citeb}{{\bf ?}\@warning
   {Citation `\@citeb' on page \thepage \space undefined}}
   {\csname b@\@citeb\endcsname}}}{#1}}

\newif\if@cghi
\def\cite{\@cghitrue\@ifnextchar [{\@tempswatrue
   \@citex}{\@tempswafalse\@citex[]}}
\def\citelow{\@cghifalse\@ifnextchar [{\@tempswatrue
   \@citex}{\@tempswafalse\@citex[]}}
\def\@cite#1#2{{$\null^{#1}$\if@tempswa\typeout
   {IJCGA warning: optional citation argument
   ignored: `#2'} \fi}}

 1
 1
 1

\font\ninerm=cmr9

\def\abstract#1{{
   \centering{\begin{minipage}{15truecm}\baselineskip=15pt\noindent
   \centerline{\bf ABSTRACT}\vspace*{3.0ex}
   \parindent=0pt #1
   \end{minipage}}\par}}
\newcommand{\nc}{\newcommand}
\nc{\rem}[1]{{\bf [#1]}}
\nc{\eqn}[1]{eq.~(\ref{#1})}
\nc{\eqns}[2]{eqs.~(\ref{#1}) and (\ref{#2})}
\nc{\fig}[1]{fig.~\ref{fig:#1}}
\nc{\figs}[2]{figs.~\ref{fig:#1} and \ref{fig:#2}}
\nc{\be}{\begin{equation}}
\nc{\ee}{\end{equation}}
\nc{\ba}{\begin{array}}
\nc{\ea}{\end{array}}
\nc{\bea}{\begin{eqnarray}}
\nc{\eea}{\end{eqnarray}}
\nc{\del}{\partial}
\nc{\eqconst}{\, \mbox{(const.)}\, }
\nc{\eqand}{\quad\mbox{and}\quad}
\nc{\eqRe}{\mbox{Re}\, }
\nc{\eqIm}{\mbox{Im}\, }
\nc{\alt}{\buildrel < \over {_\sim}}
\nc{\agt}{\buildrel > \over {_\sim}}
\nc{\st}{\scriptstyle}
\nc{\sst}{\scriptscriptstyle}
\nc{\mco}{\multicolumn}
\nc{\vep}{\varepsilon}
\nc{\lr}{\leftrightarrow}
\nc{\ra}{\rightarrow}
\nc{\vp}{{\bf p}}
\nc{\al}{\alpha}
\nc{\ab}{\bar{\alpha}}
\nc{\semi}{;\ }
\nc{\spa}[3]{\left\langle#1\,#3\right\rangle}
\nc{\spb}[3]{\left[#1\,#3\right]}
\nc{\LP}{\left(}
\nc{\RP}{\right)}
\nc{\LB}{\left[}
\nc{\RB}{\right]}
\nc{\Tr}{\mathop{\rm Tr}\nolimits}
\nc{\tr}{\mathop{\rm tr}\nolimits}
\nc{\e}{\epsilon}
\nc{\cg}{c_\Gamma}
\nc{\hf}{\textstyle{1\over2}}
\nc{\Li}{\mathop{\rm Li}\nolimits}
\nc{\Ls}{\mathop{\rm Ls}\nolimits}
\nc{\Ll}{\mathop{\rm L}\nolimits}
\nc{\gluino}{{\tilde g}}
\nc{\qb}{{\bar q}}
\nc{\SUSY}{{\rm SUSY}}
\nc{\susy}{{\rm SUSY}}
\nc{\tree}{{\rm tree}}
\nc{\oneloop}{{\rm 1-loop}}
\nc{\Atree}{A^{\rm tree}}
\nc{\Atreestar}{A^{\rm tree\,*}}
\nc{\Aloop}{A^{\rm 1-loop}}
\nc{\Aloopstar}{A^{\rm 1-loop\,*}}
\nc{\pol}{\varepsilon}
\nc{\si}{\sigma}
\nc{\ns}{n_{\mskip-2mu s}}
\nc{\nf}{n_{\mskip-2mu f}}
\nc{\ib}{{\bar\imath}}
\nc{\jb}{{\bar\jmath}}
\nc{\treemhv}{{\rm tree\ MHV}}
\nc{\loopmhv}{{\rm 1-loop\ MHV}}
\nc{\dlips}{d{\rm LIPS}}
\nc{\lsl}{\not{\hbox{\kern-2.3pt $\ell$}}}
\nc{\ksl}{\not{\hbox{\kern-2.3pt $k$}}}
\nc{\Slash}[1]{\slash\hskip -0.17 cm #1}
\nc{\tn}[2]{t^{[#1]}_{#2}}
\input epsf
\newcommand{\figdir}{}
\newcommand{\figmac}[5]{\begin{figure}
\centerline{\parbox[t]{#1in}{\epsfbox{\figdir #2.ps}}}
\caption[#4]{\label{fig:#3} #5}\end{figure}}
\def\epsfsize#1#2{\ifdim#1>\hsize\hsize\else#1\fi}
\begin{document}
\vspace{3.0cm}
\noindent hep-ph/9507214 \hfill SLAC--PUB--95--6946\break
\rightline{July, 1995}

\vspace*{3.0cm}
\centerline{\large\bf STRINGS AND SUPERSYMMETRY}
\vspace*{3.0ex}
\centerline{\large\bf AS TOOLS FOR PERTURBATIVE QCD\footnote{%
Research supported by the Department of Energy under grant
DE-AC03-76SF00515.}}

\vspace*{6.0ex}
\centerline{\large\rm LANCE DIXON}
\vspace*{1.5ex}
\centerline{\large\it Stanford Linear Accelerator Center}
\vspace*{1.0ex}
\centerline{\large\it Stanford University, Stanford, CA 94309}
\vspace*{4.5ex}
\vskip3.0truecm
\abstract{
We review techniques simplifying the analytic calculation
of one-loop QCD amplitudes with many external legs, for use in
next-to-leading-order corrections to multi-jet processes.
We explain how a supersymmetry-inspired organization
works well in conjunction with other tools, namely
the color and helicity decompositions of amplitudes,
and the constraints imposed by perturbative
unitarity and collinear singularities.
String theory seems most useful as a heuristic guide.
Using these techniques,
the complete set of one-loop five-parton QCD amplitudes, as well
as certain sequences of special helicity amplitudes with an
arbitrary number of external gluons, have been obtained.
}

\vspace{2.0cm}
\centerline{\sl Presented at SUSY95, Paris, France, May 15-19, 1995}

\vfill\eject

%%%%%%%%%%%%%%%%%%%%%%%%%%%%%%%%%%%%%%%%%%%%%%
\pagestyle{plain}
\setlength{\baselineskip}{3.0ex}
\setcounter{footnote}{0}
\renewcommand{\thefootnote}{\alph{footnote}}
%%%%%%%%%%%%%%%%%%%%%%%%%%%%%%%%%%%%%%%%%%%%%%

\section{Motivation}
Although most people at this conference may be convinced that
supersymmetry and superstring theory are realized in Nature,
let us suppose for the sake of argument that they are not.
Can these beautiful theories then have any practical implications?
The answer is yes; they can still help to organize complicated
perturbative gauge theory calculations, particularly in QCD.
We will argue that supersymmetry works best in conjunction with a
number of other tools:
the color and helicity decompositions of amplitudes,
and the constraints imposed by perturbative
unitarity and collinear singularities.
At present, at least for one-loop calculations, string theory seems
more useful as a heuristic guide to how to organize calculations,
rather than as a detailed calculation tool.

Next-to-leading-order (NLO) perturbative QCD corrections
are important for precision comparison of theoretical predictions
with collider experiments, for many multi-jet and jet-associated
processes.  Currently, NLO results are only available for processes
involving four ``partons'' (one or more of the partons may be
replaced by a $\gamma$, $Z$ or $W$), for example
$p\bar{p}\to 2$~jets\cite{NLOTwoJets}
and $e^+e^-\to 3$~jets.\cite{eeThreeJets}
There are two parts to an NLO correction to an $n$-parton process:
a real (or tree) part, obtained by integrating the tree-level
cross-section for a $(n+1)$-parton process over an ``unobserved''
portion of phase space;
and a virtual (or one-loop) part, obtained by interfering the
one-loop $n$-parton amplitude with the corresponding tree amplitude.
Because the calculation of tree amplitudes is now fairly
efficient,\cite{ManganoParke}
it is the calculation of one-loop multi-parton amplitudes
that forms the ``analytical bottleneck'' to
producing NLO results for more complicated processes.
(It is not the only obstacle, however; much numerical work is
required to combine the real and virtual corrections.)
The difficulty in going to more than four external partons
is indicated by the time lag between the
calculation of one-loop four-parton amplitudes in 1980\cite{eeThreeJets}
and 1986\cite{EllisSexton} and that of five-parton
amplitudes in the past two years.\cite{FiveGluon,Kunsztqqqqg,qqggg}

In principle it is straightforward to compute one-loop amplitudes
by drawing all Feynman diagrams and evaluating them using standard
reduction techniques for the loop integrals.  In practice this method
becomes extremely inefficient and cumbersome as the number of
external legs grows, because there are:
\par\noindent
1. {\bf too many diagrams} --- many diagrams are related by gauge
invariance, and
\par\noindent
2. {\bf too many terms in each diagram} --- nonabelian gauge boson
self-interactions are complicated.
\par\noindent
Consequently, intermediate expressions tend to be vastly more
complicated than the final results, when the latter are
represented in an appropriate way.

A useful organizational framework, that helps tame the size of
intermediate expressions, is Total Quantum-number Management
(TQM), which suggests to:
\par\noindent
$\bullet$ Keep track of all quantum numbers of external
particles --- namely, {\it helicity} and {\it color} information.
\par\noindent
$\bullet$ Use the helicity/color information to decompose the amplitude
into simpler, gauge-invariant pieces, called
{\it primitive amplitudes}.
\par\noindent
$\bullet$ Use supersymmetry to organize the sum over {\it internal}
particle spins in the loop.
\par\noindent
$\bullet$ Square amplitudes to get probabilities, and sum over
helicities and colors to obtain unpolarized cross-sections,
only at the very {\it end} of the calculation.
\par\noindent
Carrying out the last step explicitly would generate a large
analytic expression; however, at this stage one would
typically make the transition to numerical evaluation,
in order to combine the virtual and real corrections.
The use of TQM is hardly new, particularly in tree-level
applications\cite{ManganoParke} ---
but it is especially useful at loop level.

%%%%%%%%%%%%%%%%%%%%%%%%%%%%%%%%%%%%%%%%%%%%%%

\section{What about string theory?}
What role can superstring theory play in this approach?
String theory has the advantage that all field theory diagrams
at a given order of perturbation theory are lumped into a single
string diagram.  Circulating in the loop(s) of the diagram are
not just the quarks and gluons of QCD, but the entire massive tower
of string excitations, with masses of order the Planck mass.
In the experimentally relevant low-energy limit, the massive states
decouple, and the string diagram degenerates into a number of
field-theory-like diagrams, determined by simple
(Bern-Kosower\cite{StringBased}) rules that are nicely compatible
with the color and helicity decompositions.

However, there are also a few disadvantages to a direct string-based
approach.  Simple rules can still generate a large mess in
intermediate steps.  (This was found to be the case in the calculation
of one-loop five-gluon amplitudes.)
The Bern-Kosower rules were derived for one-loop amplitudes
with external gluons only, and new rules would have to be rederived
for external quarks and/or multi-loop amplitudes.
Some progress has been made in these
directions,\cite{PasqRol} but not yet to the
point of pushing the field theory state-of-the-art.
On the other hand, string-based rules can be mimicked in field
theory\cite{Mapping} by a combination of
background-field\cite{Background} and Gervais-Neveu\cite{GN} gauges.
Such gauge choices can be used with external fermions too.
Finally, other tools --- in particular supersymmetry, unitarity and
collinear limits --- can be even more efficient routes to one-loop
scattering amplitudes.  String theory remains useful as a heuristic
guide; we will give a couple of examples below.

%%%%%%%%%%%%%%%%%%%%%%%%%%%%%%%%%%%%%%%%%%%%%%

\section{Color and helicity decomposition}
As an example of the color and helicity decomposition of a one-loop
QCD amplitude, consider the amplitude for $n$ external
gluons, all taken to be outgoing.
We generalize the $SU(3)$ color group to
$SU(N_c)$, and label the gluons $i=1,2,\ldots,n$ by their adjoint
color indices $a_i=1,2,\ldots,N_c^2-1$, and helicities
$\lambda_i = \pm$.  Without giving the details of the helicity
decomposition formalism, it is convenient to use gluon circular
polarization vectors expressed in terms of
massless Weyl spinors.\cite{SpinorHelicity}
The color decomposition\cite{LoopColor} is performed in
terms of traces of $SU(N_c)$ generators $T^a$ in the
fundamental representation, with $\Tr(T^aT^b)=\delta^{ab}$,
\bea \label{ColorDecomp}
&&\hskip -8mm
 {\cal A}^\oneloop_n \LP \{k_i,\lambda_i,a_i\}\RP =
  g^n\Biggl[
    \sum_{\sigma \in S_n/Z_n}
    N_c\,\Tr\LP T^{a_{\sigma(1)}}\cdots T^{a_{\sigma(n)}}\RP\
     A_{n;1}(\sigma(1^{\lambda_1}),\ldots,\sigma(n^{\lambda_n}))
     \nonumber \\
&& \hskip -8mm
 +\ \sum_{c=2}^{\lfloor{n/2}\rfloor+1}
      \sum_{\sigma \in S_n/S_{n;c}}
    \Tr\LP T^{a_{\sigma(1)}}\cdots T^{a_{\sigma(c-1)}}\RP\
    \Tr\LP T^{a_{\sigma(c)}}\cdots T^{a_{\sigma(n)}}\RP\
     \ A_{n;c}(\sigma(1^{\lambda_1}),\ldots,\sigma(n^{\lambda_n}))
     \Biggr]\, ,
\eea
where $A_{n;c}$ are the partial amplitudes, $g$ is the gauge coupling,
and $\lfloor{x}\rfloor$ is the integer part of $x$.
$S_n$ is the set of all permutations of $n$ objects,
while $Z_n$ and $S_{n;c}$ are the subsets of $S_n$ that leave
the corresponding single and double trace structures invariant.

The color decomposition~(\ref{ColorDecomp})
can be derived fairly simply from Feynman
diagrams in the double-line formalism;\cite{DoubleLine}
however, it is even more transparent to derive it from open string
theory.   The end of an open string carries the color information ---
an index $i$ in the fundamental $N_c$ representation.
A vertex operator for an external gluon with adjoint index $a$
carries the Chan-Paton factor\cite{ChanPaton} $(T^a)_i^{~\jb}$
and rotates the fundamental index.
At tree-level, the string world-sheet is a topologically a disk,
and the Chan-Paton factors hook together into a single
trace of the form $\Tr\LP T^{a_{1}}\cdots T^{a_{n}}\RP$
(or some permutation thereof).  At one-loop, the (oriented) world-sheet
is an annulus, and there are two possibilities, shown in
fig.~\ref{fig:openstring}:
either (a) all vertices attach to the same edge of the annulus,
or (b) some vertices attach to each of the two edges.
In case (a) one obtains the single-trace structure
of the first term in \eqn{ColorDecomp}, which multiplies $A_{n;1}$;
the factor of $N_c = \Tr(1)$
comes from the edge with no vertices attached to it.
In the second case one obtains the remaining double-trace terms
in~(\ref{ColorDecomp}), which multiply $A_{n;c>1}$.

%=== figure =====
\figmac{3.5}{opencolor}{openstring}{
The two types of oriented one-loop open string world-sheet diagrams:
Either (a) all vertices attach to the same edge of the annulus,
or (b) some vertices attach to each edge}
{The two types of oriented one-loop open string world-sheet diagrams:
Either (a) all vertices attach to the same edge of the annulus,
or (b) some vertices attach to each edge.}

Closer inspection of the two different types of string world-sheet
diagrams in the low-energy limit {\it suggests} a possible relation
between the corresponding $A_{n;c>1}$ and $A_{n;1}$ amplitudes.
Apart from the fact that some vertex operators
appear on the ``wrong side'' of the annulus --- which one might hope is
irrelevant in the low-energy limit --- the $A_{n;c}$ world-sheet diagrams
appear to be just the sum over a particular set $COP\{\alpha\}\{\beta\}$
of permutations of the $A_{n;1}$ diagrams,
those that preserve the cyclic ordering of the sets $\{\alpha\}$ and
$\{\beta\}$ of vertices on the respective inner and outer boundaries
of the annulus.
Thus the following formula is suggested,
\be \label{Anc}
 A_{n;c}(1,2,\ldots,c-1;c,c+1,\ldots,n)\ =\
 (-1)^{c-1} \sum_{\sigma\in COP\{\alpha\}\{\beta\}} A_{n;1}(\sigma),
\ee
where $\{\alpha\} \equiv \{c-1,c-2,\ldots,2,1\}$,
$\{\beta\} \equiv \{c,c+1,\ldots,n-1,n\}$.
This formula is of practical use because now only the $A_{n;1}$ have to
be calculated.
It can be proven using Feynman diagrams,\cite{SusyFour}
but its motivation from the structure of open string theory is a good
example of how strings can serve as a heuristic guide to gauge theory
organization.

Equation~(\ref{Anc}) shows that
the $A_{n;1}$ are the more basic objects, so we call them
{\it primitive amplitudes}.  They are ``color-ordered'' amplitudes, in
that they only receive contributions from diagrams with a
particular cyclic ordering of the gluons around the loop.
This greatly simplifies their analytic structure, because cuts and
poles can only appear in channels formed by the sum of
{\it cyclically adjacent} momenta,
$(k_i+k_{i+1}+\cdots+k_{i+r-1})^2$.

Even the $A_{n;1}$ are not all independent,
due to parity and cyclic invariance.
For example, for $n=5$ only four are independent,
$A_{5;1}(1^+,2^+,3^+,4^+,5^+)$,
$A_{5;1}(1^-,2^+,3^+,4^+,5^+)$,
$A_{5;1}(1^-,2^-,3^+,4^+,5^+)$, and
$A_{5;1}(1^-,2^+,3^-,4^+,5^+)$.
The first two are not required at NLO because the corresponding
tree helicity amplitudes vanish, and are very simple for the
same reason.
Analytic expressions for the latter two are more
complex\cite{FiveGluon} but still ``fit on a page'' (see below).
In contrast, the color- and helicity-summed virtual correction to the
cross-section, built from permutation sums of the latter two primitive
amplitudes, would fill hundreds of pages.

%%%%%%%%%%%%%%%%%%%%%%%%%%%%%%%%%%%%%%%%%%%%%%

\section{What about supersymmetry?}
Supersymmetry plays a role even in a non-supersymmetric theory such as
QCD.  This is because tree-level QCD is ``effectively''
supersymmetric.\cite{NewSWI}   Consider the $n$-gluon tree amplitude.
It has no loops in it, therefore it has no fermion loops in it.
Therefore the fermions in the theory might as well be gluinos, i.e.
at tree-level the theory might as well be super Yang-Mills theory.
The ``non-supersymmetry'' of QCD only leaks in at the loop level.

Supersymmetric results are often simpler than non-supersymmetric ones.
For example, the anomalous magnetic moments of the electron in QED
and in super-QED are\cite{gQEDref,gSQEDref}
\bea
&&\hskip-0.7cm
{g_e-2\over2}\Biggr\vert_{\rm QED} =
  {\alpha\over2\pi}
+ \left[  {197\over144}    + {\pi^2\over12}
        + {3\zeta_3\over4} - {\pi^2\ln2 \over 2} \right]
    \left({\alpha\over\pi}\right)^2
+ 1.1761(4) \left({\alpha\over\pi}\right)^3
- 1.43(14)  \left({\alpha\over\pi}\right)^4 + \cdots
\nonumber \\
\label{gmtwo}
&&\hskip-0.8cm
{g_e-2\over2}\Biggr\vert_{\rm SQED} = 0;
\eea
the latter is an example of a supersymmetry Ward identity
(SWI).\cite{gSQEDref}
Here the supersymmetric result is ``too simple'':
it does not form a significant part of the non-supersymmetric result.
For one-loop multi-parton QCD calculations, the situation is somewhat
more favorable.

Supersymmetry Ward identities\cite{SWI} can be derived for general
$S$-matrix elements $\langle \Phi_1 \cdots \Phi_n \rangle$ using
the fact that the supercharge $Q$ annihilates
the vacuum; when the fields $\Phi_i$ create helicity eigenstates,
many of the $[Q,\Phi_i]$ terms can be arranged to vanish.
Taking all particles to be outgoing, the
simplest identities are for amplitudes with at most two negative
helicities, and the rest positive\cite{SWI,NewSWI,ManganoParke}:
\bea \label{vanishSWI}
  A_n^\SUSY(1^\pm,2^+,3^+,\ldots,n^+)\ &=&\ 0, \\
 \label{MHVSWI}
 A_n^\SUSY(1^-,2_P^-,3_P^+,4^+,\ldots,n^+)
\ &=&\ \left({\spa1.2\over\spa1.3}\right)^{2|h_P|}
  A_n^\SUSY(1^-,2_\phi^-,3_\phi^+,4^+,\ldots,n^+).
\eea
Here $\phi$ stands for a scalar particle (for which the ``helicity''
$\pm$ means particle vs. antiparticle), while $P$ stands for a scalar,
fermion or gluon, with respective helicity $h_P = 0, \hf, 1$.
We have introduced spinor product
notation,\cite{SpinorHelicity,ManganoParke}
$\spa{j}.{l} = \langle j^- | l^+ \rangle = \bar{u}_-(k_j) u_+(k_l)$
and
$\spb{j}.{l} = \langle j^+ | l^- \rangle = \bar{u}_+(k_j) u_-(k_l)$,
where $u_\pm(k)$ is a massless Weyl spinor with momentum $k$ and
chirality $\pm$.

The SWI hold order-by-order in perturbation theory.
They apply directly to all {\it tree-level} QCD amplitudes
because of the ``effective'' supersymmetry described above.
They guide the simple structure of ``maximally helicity violating''
(MHV) QCD tree amplitudes, which for $n$ external gluons
are\cite{ParkeTaylor,BerendsGiele}
\bea \label{vanishQCD}
  A_n^\tree(1^\pm,2^+,3^+,\ldots,n^+)\ &=&\ 0, \\
  A_n^\tree(1^+,\ldots,j^-,\ldots,k^-,\ldots,n^+)\ &=&\
 \label{MHVQCD}
  i\,{\spa{j}.{k}^4 \over \spa1.2 \cdots \spa{n}.1}\ .
\eea
Even in the second, nonvanishing case, the amplitude
remains simple because the SWI forbid the appearance of multi-particle
poles (poles in $(k_i+\cdots+k_{i+r-1})^2$ with $r>2$).
The intermediate gluon in the factorization of \eqn{MHVQCD} on a
multi-particle pole has negative helicity as seen by one of the two
lower-point amplitudes, but positive helicity as seen by the other.
Thus the two lower-point amplitudes share a total of three negative
helicities, and so one of them must vanish by \eqn{vanishSWI}.

At loop level, QCD ``knows'' that it is not supersymmetric.
However, one can use supersymmetry to trade gluons in the loop
diagrams for scalars.   Scalars lead to algebraically simpler
diagrams, because they cannot propagate spin information around
the loop.  For an amplitude with all external gluons,
we rewrite the internal gluon loop $g$ (and fermion loop $f$) as a
supersymmetric contribution plus a complex scalar loop $s$,
\bea \label{SusyDecomp}
  g &=& (g+4f+3s)\ -\ 4(f+s)\ +\ s\ =\ A^{N=4}\ -\ 4\,A^{N=1}
  \ +\ A^{\rm scalar},
  \nonumber \\
  f &=& (f+s)\ -\ s\ =\ A^{N=1}\ -\ A^{\rm scalar},
\eea
where $A^{N=4}$ represents the contribution of the $N=4$ super Yang-Mills
multiplet, and $A^{N=1}$ an $N=1$ chiral matter supermultiplet.
In the context of TQM, this use of supersymmetry could be termed
``internal spin management''.

As an example, let's look at the five-gluon primitive
amplitude $A_{5;1}(1^-,2^-,3^+,4^+,5^+)$,
whose components according to~(\ref{SusyDecomp}) are\cite{FiveGluon}
\bea \label{gggggmmppp}
  A^{N=4}\ &=&\ \cg \, A^\tree \sum_{j=1}^5 \Biggl[
     -{1\over\e^2} \LP {\mu^2\over -s_{j,j+1}}\RP^\e
     + \ln\LP{-s_{j,j+1}\over -s_{j+1,j+2}}\RP\,
       \ln\LP{-s_{j+2,j-2}\over -s_{j-2,j-1}}\RP
     + {\pi^2\over6} \Biggr]
\nonumber \\
  A^{N=1}\ &=&\ \cg \, A^\tree \Biggl[ {5\over2\e}
    +{1\over2}\LB\ln\LP{\mu^2\over -s_{23}}\RP
                +\ln\LP{\mu^2\over -s_{51}}\RP\RB + 2 \Biggr]
\nonumber \\
&&\quad + {i\cg\over2}
   {{\spa1.2}^2 \LP\spa2.3\spb3.4\spa4.1+\spa2.4\spb4.5\spa5.1\RP\over
    \spa2.3\spa3.4\spa4.5\spa5.1}
     {\ln\LP {-s_{23}\over -s_{51}}\RP\over s_{51}-s_{23}}
\nonumber \\
  A^{\rm scalar}\ &=&\ {1\over3} A^{\rm N=1}
    + {2\over9} \cg \, A^\tree
\nonumber \\
&&\hskip-3mm + {i\cg\over3} \Biggl[
   - { \spb3.4\spa4.1\spa2.4\spb4.5
       \LP\spa2.3\spb3.4\spa4.1+\spa2.4\spb4.5\spa5.1\RP
          \over\spa3.4\spa4.5 }
     { \ln\LP{-s_{23}\over -s_{51}}\RP
         -{1\over2}\LP{s_{23}\over s_{51}}-{s_{51}\over s_{23}}\RP
               \over (s_{51}-s_{23})^3 }
\nonumber \\
&& \hskip-0mm
   - {\spa3.5{\spb3.5}^3\over\spb1.2\spb2.3\spa3.4\spa4.5\spb5.1}
   + {\spa1.2{\spb3.5}^2\over\spb2.3\spa3.4\spa4.5\spb5.1}
   + {1\over2}{\spa1.2\spb3.4\spa4.1\spa2.4\spb4.5\over
                  s_{23}\spa3.4\spa4.5 s_{51}} \Biggr]\ ,
\eea
where $A^\tree = A_5^\tree(1^-,2^-,3^+,4^+,5^+)$ is given
in \eqn{MHVQCD}, and
\be \label{cgammadef}
  c_\Gamma = {\Gamma(1+\e)\Gamma^2(1-\e)
         \over(4\pi)^{2-\e}\Gamma(1-2\e)}\ .
\ee
We see that the three components have quite different analytic
structure, indicating that the rearrangement~(\ref{SusyDecomp}) is
a natural one.  The $N=4$ supersymmetric component is the
simplest, followed by the $N=1$ chiral component.
The non-supersymmetric scalar component is the most complicated,
and the hardest to calculate.
Yet it is still simpler than the direct gluon calculation,
because it does not mix all three components together.

%%%%%%%%%%%%%%%%%%%%%%%%%%%%%%%%%%%%%%%%%%%%%%

\section{One-loop amplitudes via unitarity}
The absorptive parts (cuts) of loop amplitudes can be determined
from phase-space integrals of products of lower-order amplitudes,
exploiting the perturbative unitarity of the $S$-matrix.
For one-loop multi-parton amplitudes, there are several reasons why
this calculation of the cuts is much easier than a direct loop
calculation:
\par\noindent
$\bullet$ One can simplify the tree amplitudes {\it before} feeding
them into the cut calculation.
\par\noindent
$\bullet$ The tree amplitudes are usually quite simple, because they
possess ``effective'' supersymmetry, even if the full loop amplitudes
do not.
\par\noindent
$\bullet$ One can further use on-shell conditions for the intermediate
legs in evaluating the cuts.

The catch is that it is not always possible to reconstruct the
full loop amplitude from its cuts.  In general there can be an
additive ``polynomial ambiguity'' --- in addition to the usual
logarithms and dilogarithms of loop amplitudes, there may be
polynomials (actually rational functions) in the kinematic variables,
which cannot be detected by the cuts.
This ambiguity is absent in one-loop massless supersymmetric
amplitudes,\cite{SusyFour,SusyOne} because of their better ultraviolet
behavior.  Notice that in the five-gluon example~(\ref{gggggmmppp})
all the polynomial terms are intimately linked to the logarithms in
both $A^{N=4}$ and $A^{N=1}$, while they are not linked in
$A^{\rm scalar}$.

To see the supersymmetric cancellations for $n$-gluon amplitudes,
it suffices to use the second-order formalism for the fermion loop,
and background-field gauge\cite{Background} for the gluon loop,
in the effective action $\Gamma(A)$.
The scalar, fermion and gluon contributions are
\bea \label{EffectiveAction}
\Gamma^{\rm scalar}(A) &\sim&
   + \ln\det\LP D^2 \RP, \nonumber \\
\Gamma^{\rm fermion}(A) &\sim&
   - \ln\det\LP D^2-\hf\sigma^{\mu\nu}F_{\mu\nu} \RP, \nonumber \\
\Gamma^{\rm gluon}(A) &\sim&
   + \ln\det\LP D^2-\Sigma^{\mu\nu}F_{\mu\nu} \RP,
\eea
where $D$ is the covariant derivative, $F$ is the external field
strength, and $\hf\sigma_{\mu\nu}$ ($\Sigma_{\mu\nu}$) is the
spin-${1\over2}$ (spin-1) Lorentz generator.
The leading behavior for large loop-momentum $\ell$
comes from the $D^2$ term in each case
($F_{\mu\nu}$ contains derivatives only with respect to
the external momenta).  This term cancels between the scalar and
fermion, and between the fermion and gluon in~\eqn{EffectiveAction},
hence it cancels in any supersymmetric linear combination.
The cancellation for an $m$-point graph is from $\ell^m$ down to
$\ell^{m-2}$
(since $\Tr\sigma_{\mu\nu} = \Tr\Sigma_{\mu\nu} = 0$).
It can be shown that an amplitude having this property in all graphs
can be uniquely reconstructed from its cuts.\cite{SusyOne}

The Neveu-Schwarz-Ramond representation of superstring theory, although
not manifestly supersymmetric, corresponds to both a second-order
fermion and background-field gauge formalism in field theory.
This is a second example of string theory as a heuristic guide.

Non-supersymmetric amplitudes generally cannot be directly
reconstructed from their unitarity cuts.  We did not discuss the
collinear behavior of loop amplitudes
here,\cite{AllPlus,SusyFour,BernChalmers} but they are a useful
and powerful practical tool for fixing the polynomial ambiguities,
recursively in $n$, by requiring consistent collinear factorization
in all channels.  The only drawback is the current lack of a theorem
that would guarantee the uniqueness of a polynomial expression obeying
all collinear consistency checks.
But no counterexamples are known either, for $n>5$.
It is also possible to extract ${\cal O}(\e^0)$ polynomial terms
from cuts evaluated to ${\cal O}(\e)$ in dimensional
regularization, but this is significantly harder than evaluation
of the cuts to ${\cal O}(\e^0)$.

As an example of how simple one-loop multi-parton cuts can be,
we outline here the evaluation of the cuts for an infinite
sequence of $n$-gluon amplitudes, the MHV
amplitudes in $N=4$ super-Yang-Mills theory.\cite{SusyFour}
We consider the case where the two negative helicity gluons lie
on the same side of the cut, as shown in fig.~\ref{fig:casea}.
(The case where they lie on the opposite side of the cut
can be quickly reduced to this case\cite{SusyFour}
 using the SWI~(\ref{MHVSWI}).)
Contributions to this cut from intermediate fermions or scalars
vanish using the ``effective'' supersymmetry of the tree amplitudes,
\eqn{vanishSWI}, and the conservation of fermion helicity and
scalar particle number.
The only contribution is from intermediate gluons
with the helicity assignment shown in fig.~\ref{fig:casea}.
The tree amplitudes on either side of the cut are
pure-glue MHV tree amplitudes, given in \eqn{MHVQCD}.

%=== figure =====
\figmac{4}{samesidecut}{casea}{
The possible intermediate helicities when both
negative helicity gluons lie on the same side of the cut}
{The possible intermediate helicities when both
negative helicity gluons lie on the same side of the cut.}

Let $j$ and $k$ denote the negative helicity external gluons.
The cut for this MHV loop amplitude, $A_{jk}^\loopmhv(1,2,\ldots,n)$,
in the channel $(k_{m_1}+k_{m_1+1}+\cdots+k_{m_2-1}+k_{m_2})^2$,
where $m_1 \leq k < j \leq m_2$, is then given by
\bea \label{caseacut}
  &&\int \dlips(-\ell_1,\ell_2)
    A^\treemhv_{jk}(-\ell_1,m_1,\ldots,m_2,\ell_2)
  \ A^\treemhv_{(-\ell_2)\ell_1}(-\ell_2,m_2+1,\ldots,m_1-1,\ell_1)
 \nonumber \\
  &=& i \, A_{jk}^\treemhv(1,2,\ldots,n)
\int \dlips(-\ell_1,\ell_2)
   { \spa{(m_1-1)}.{m_1} \spa{\ell_1}.{\ell_2}
    \over \spa{(m_1-1)}.{\ell_1} \spa{\ell_1}.{m_1} }
   { \spa{m_2}.{(m_2+1)} \spa{\ell_2}.{\ell_1}
    \over \spa{m_2}.{\ell_2} \spa{\ell_2}.{(m_2+1)} }\ ,
\eea
where the spinor products are labelled by either loop
momenta ($\ell_1$, $\ell_2$) or external particle labels.
The $(4-2\e)$-dimensional
Lorentz-invariant phase space measure is denoted by
$\dlips(-\ell_1,\ell_2)$.
The cut~(\ref{caseacut}) remains simple for arbitrarily many external
gluons, thanks to the simple form of the MHV tree
amplitudes~(\ref{MHVQCD}) --- $\ell_1$ and $\ell_2$
appear in only a few of the factors.

The integral~(\ref{caseacut}) can be viewed as a cut hexagon loop integral.
To see this, use the on-shell condition $\ell_1^2=\ell_2^2=0$
to rewrite the four spinor product denominators in~(\ref{caseacut})
as scalar propagators, multiplied by a numerator factor.  For
example,
$1/\spa{\ell_1}.{m_1}
= \spb {m_1}.{\ell_1} / ( \spa{\ell_1}.{m_1} \spb {m_1}.{\ell_1} )
= \spb {m_1}.{\ell_1} / ( 2 \ell_1 \cdot k_{m_1} )
= -\spb {m_1}.{\ell_1} / (\ell_1 - k_{m_1})^2$.
In addition to these four propagators, there are two cut propagators
implicit in the phase-space integral $\int \dlips(-\ell_1,\ell_2)$.
The Schouten identity,
$\spa a.b\spa c.d = \spa a.d\spa c.b + \spa a. c\spa b.d$,
lets us rewrite the integrand of~(\ref{caseacut}) as
\bea \label{firstpartialfr}
 {\cal I}\ &=&\
   \biggl( { \spa{(m_1-1)}.{\ell_2} \over \spa{(m_1-1)}.{\ell_1} }
         - { \spa{m_1}.{\ell_2} \over \spa{m_1}.{\ell_1} } \biggr)
   \biggl( { \spa{m_2}.{\ell_1} \over \spa{m_2}.{\ell_2} }
         - { \spa{(m_2+1)}.{\ell_1} \over \spa{(m_2+1)}.{\ell_2} }
   \biggr)
 \nonumber \\
 &=&\ -{\spa{m_1}.{\ell_2}\spa{m_2}.{\ell_1} \over
      \spa{m_1}.{\ell_1}\spa{m_2}.{\ell_2}}
   \pm \Bigl[\ m_1 \lr (m_1-1)\ ,\ m_2 \lr (m_2+1)\ \Bigr]\ ,
\eea
antisymmetrizing in each exchange.  In terms of propagators,
\bea \label{traceterms}
 {\cal I}\ &=&\
    {\spb{\ell_1}.{m_1}\spa{m_1}.{\ell_2}
     \spb{\ell_2}.{m_2}\spa{m_2}.{\ell_1}
    \over (\ell_1-k_{m_1})^2 (\ell_2+k_{m_2})^2}
    \pm \Bigl[\ m_1 \lr (m_1-1)\ ,\ m_2 \lr (m_2+1)\ \Bigr]
 \nonumber \\
 &=&\ {\tr_+(\lsl_1 \ksl_{m_1} \lsl_2 \ksl_{m_2})
     \over (\ell_1-k_{m_1})^2 (\ell_2+k_{m_2})^2}
    \pm \Bigl[\ m_1 \lr (m_1-1)\ ,\ m_2 \lr (m_2+1)\ \Bigr]\ ,
\eea
where the $\tr_+$
indicates the insertion of a $(1+\gamma_5)/2$ projector into the
trace.  Thus we have reduced the cut hexagon integral~(\ref{caseacut})
to a sum of four cut box integrals.

A straightforward Passarino-Veltman reduction\cite{PV} expresses the
box integrals from~(\ref{traceterms}) in terms of scalar boxes,
triangles and bubbles.
The coefficients of the triangles and bubbles vanish.
The only scalar boxes with non-vanishing coefficients are
those with two diagonally opposite massless legs.
The full amplitude, which matches the cuts in all channels, is
\be \label{MHVanswer}
  A_{n;1}^{N=4}(1^+,\ldots,j^-,\ldots,k^-,\ldots,n^+)\ =\
 i \, \cg \, \mu^{2\e} \,
   { \spa{j}.k^4 \over \spa1.2\spa2.3\cdots\spa{n}.1 }\, V_n\,,
\ee
where $V_n$ is a universal, cyclically symmetric function,
\be
  V_{2m+1} = \sum_{r=1}^{m-1} \sum_{i=1}^n f_{i,r}\ ,
 \qquad\qquad\qquad
    V_{2m} = \sum_{r=1}^{m-2} \sum_{i=1}^n f_{i,r}
                  + \sum_{i=1}^{n/2} f_{i,m-1}\ .
\ee
The scalar box integral functions in $4-2\e$ dimensions are given
through ${\cal O}(\e^0)$ by
\bea \label{fexplicit}
&&\hskip-1.3cm
f_{i,r} =
- {1\over\e^2} \Bigl[ (-\tn{r+1}{i-1})^{-\e} + (-\tn{r+1}{i})^{-\e}
              - (-\tn{r}{i} )^{-\e} - (-\tn{n-r-2}{i+r+1} )^{-\e} \Bigr]
- \Li_2\hskip-0.1cm\left( \hskip-0.05cm 1 -
     {\tn{r}{i}\tn{n-r-2}{i+r+1}
                \over\tn{r+1}{i-1}\tn{r+1}{i}} \hskip-0.05cm \right)
 \nonumber \\
&&\hskip-1.4cm
+ \Li_2\hskip-0.1cm\left( \hskip-0.05cm 1 -
     {\tn{r}{i} \over \tn{r+1}{i-1}} \hskip-0.05cm \right)
\hskip-0.05cm
+ \Li_2\hskip-0.1cm\left( \hskip-0.05cm 1 -
     {\tn{r}{i} \over \tn{r+1}{i}} \hskip-0.05cm \right)
\hskip-0.05cm
+ \Li_2\hskip-0.1cm\left( \hskip-0.05cm 1 -
     {\tn{n-r-2}{i+r+1} \over \tn{r+1}{i-1}} \hskip-0.05cm \right)
\hskip-0.05cm
+ \Li_2\hskip-0.1cm\left( \hskip-0.05cm 1 -
     {\tn{n-r-2}{i+r+1} \over \tn{r+1}{i}} \hskip-0.05cm \right)
\hskip-0.05cm
+ {1\over 2} \ln^2\hskip-0.1cm\left( \hskip-0.05cm
     {\tn{r+1}{i-1} \over \tn{r+1}{i}} \hskip-0.05cm \right)
\eea
where $\tn{r}{i} = (k_i+k_{i+1}+\cdots+k_{i+r-1})^2$.
Since $\tn{1}{i} = k_i^2 = 0$, we set $(-\tn{1}{i})^{-\e}=0$
in~(\ref{fexplicit}).

%%%%%%%%%%%%%%%%%%%%%%%%%%%%%%%%%%%%%%%%%%%%%%

\section{Conclusions}
We have argued that the use of supersymmetry and string theory (the
latter more heuristically), in combination with more conventional tools
such as helicity and color decompositions, unitarity and collinear
limits, can lead to many simplifications in the calculation of one-loop
multi-parton amplitudes.   At the practical level, some of these tools
have been instrumental in calculating the one-loop five-parton
amplitudes ($ggggg$, $\qb q\qb qg$ and $\qb qggg$) which form the
analytical bottleneck to NLO cross-sections for three-jet events at
hadron colliders.\cite{FiveGluon,Kunsztqqqqg,qqggg}
They have also been used to obtain infinite sequences of special
one-loop helicity amplitudes in closed
form.\cite{AllPlus,Mahlon,SusyFour,SusyOne}
The polynomial ambiguities in the non-supersymmetric components
of one-loop QCD amplitudes are the main obstacle to their
efficient evaluation.
If one can show that these ambiguities may be fixed uniquely
(and efficiently!) using factorization limits, then this obstacle would
be lifted, and one would have a general technique for constructing
one-loop QCD amplitudes without ever evaluating genuine loop
diagrams.

%%%%%%%%%%%%%%%%%%%%%%%%%%%%%%%%%%%%%%%%%%%%%%


\section{References}
\begin{thebibliography}{9}
\bibitem{NLOTwoJets} S.D.\ Ellis, Z. Kunszt and D.E.\ Soper,
         Phys.\ Rev.\ D40:2188 (1989);
         Phys.\ Rev.\ Lett.\ 64:2121 (1990);
         Phys.\ Rev.\ Lett.\ 69:1496 (1992)\semi
         F. Aversa, M. Greco, P. Chiappetta and J.P.\
         Guillet, Phys.\ Rev.\ Lett.\ 65:401 (1990)\semi
         F. Aversa, L. Gonzales, M. Greco, P. Chiappetta
         and J.P.\ Guillet, Z. Phys.\ C49:459 (1991)\semi
         W.T. Giele, E.W.N. Glover and D.A. Kosower,
         Phys.\ Rev.\ Lett. 73:2019 (1994);
         Phys.\ Lett.\ B339:181 (1994).
\bibitem{eeThreeJets} R.K. Ellis, D.A. Ross and A.E. Terrano,
         Phys.\ Rev.\ Lett. 45:1226 (1980);
         Nucl.\ Phys.\ B178:421 (1981)\semi
         K. Fabricius, I. Schmitt, G. Kramer and G. Schierholz,
         Phys.\ Lett.\ B97:431 (1980);
         Z.\ Phys.\ C11:315 (1981).
\bibitem{ManganoParke} M. Mangano and S. Parke, Phys. Rep. 200:301
         (1991).
\bibitem{EllisSexton} R.K. Ellis and J.C. Sexton,
         Nucl.\ Phys.\ B269:445 (1986).
\bibitem{FiveGluon}  Z. Bern, L. Dixon and D.A. Kosower, Phys. Rev. Lett.
         70:2677 (1993).
\bibitem{Kunsztqqqqg} Z. Kunszt, A. Signer and Z. Tr\'ocs\'anyi,
         Phys.\ Lett.\ B336:529 (1994).
\bibitem{qqggg} Z. Bern, L. Dixon and D.A. Kosower,
         Nucl. Phys. B437:259 (1995).
\bibitem{StringBased} Z. Bern and D.A. Kosower, Phys. Rev. Lett. 66:1669
         (1991); Nucl. Phys. B379:451 (1992).
\bibitem{PasqRol} K. Roland, Phys.\ Lett.\ B289:148 (1992)\semi
         A. Pasquinucci and K. Roland, Nucl.\ Phys.\ B440:441 (1995).
\bibitem{Mapping} Z. Bern and D.C.\ Dunbar,
         Nucl.\ Phys.\ B379:562 (1992).
\bibitem{Background} G. 't Hooft, in Acta Universitatis Wratislavensis
         no.\ 38, 12th Winter School of Theoretical Physics in Karpacz,
         {\it Functional and Probabilistic Methods in Quantum Field
         Theory}, Vol. 1 (1975)\semi
         B.S.\ DeWitt, in {\it Quantum gravity II}, eds. C. Isham,
         R.\ Penrose and D.\ Sciama (Oxford, 1981)\semi
         L.F.\ Abbott, Nucl.\ Phys.\ B185:189 (1981)\semi
         L.F. Abbott, M.T. Grisaru and R.K. Schaefer,
         Nucl.\ Phys.\ B229:372 (1983).
\bibitem{GN} J.L.\ Gervais and A. Neveu, Nucl.\ Phys.\ B46:381 (1972).
\bibitem{SpinorHelicity} F.A. Berends, R. Kleiss, P. De Causmaecker,
         R. Gastmans and T.T. Wu, Phys. Lett. B103:124 (1981);
         P. De Causmaeker, R. Gastmans,  W. Troost and T.T. Wu,
         Nucl. Phys. B206:53 (1982);
         R. Kleiss and W.J. Stirling, Nucl. Phys. B262:235 (1985);
         J.F. Gunion and Z. Kunszt, Phys. Lett. B161:333 (1985);
         Z. Xu, D.-H.\ Zhang and L. Chang, Nucl. Phys. B291:392 (1987).
\bibitem{LoopColor}  Z. Bern and D.A. Kosower, Nucl. Phys. B362:389
         (1991).
\bibitem{DoubleLine} G. 't Hooft, Nucl. Phys. B72:461 (1974);
         Nucl. Phys. B75:461 (1974)\semi
         see also P.~Cvitanovi\'c, {\it Group Theory} (Nordita, 1984).
\bibitem{ChanPaton} J. Paton and H.-M. Chan, Nucl.\ Phys.\ B10:519
         (1969).
\bibitem{SusyFour} Z. Bern, D. Dunbar, L. Dixon and D.A. Kosower,
         Nucl. Phys. B425:217 (1994).
\bibitem{NewSWI} S.J. Parke and T. Taylor, Phys.\ Lett.\ B157:81 (1985)\semi
         Z. Kunszt, Nucl.\ Phys.\ B271:333 (1986).
\bibitem{gQEDref} J. Schwinger, Phys.\ Rev.\ 73:416 (1948)\semi
         R. Karplus and N.M. Kroll, Phys.\ Rev.\ 77:536 (1950)\semi
         A. Petermann, Helv.\ Phys.\ Acta 30:407 (1957)\semi
         C. Sommerfield, Phys.\ Rev.\ 107:328 (1957)\semi
         P. Cvitanovi\'c and T. Kinoshita, Phys.\ Rev.\ D10:4007
         (1974)\semi
         T. Kinoshita and W.B. Lindquist, Phys.\ Rev.\ Lett.\ 47:1573
         (1981).
\bibitem{gSQEDref} S. Ferrara and E. Remiddi, Phys.\ Lett.\ B53:347
         (1974)\semi
         S. Ferrara and M. Porrati, Phys.\ Lett.\ B288:85 (1992).
\bibitem{SWI} M.T.\ Grisaru, H.N.\ Pendleton and P.\ van Nieuwenhuizen,
         Phys. Rev. {D15}:996 (1977)\semi
         M.T. Grisaru and H.N. Pendleton, Nucl.\ Phys.\ B124:81 (1977).
\bibitem {ParkeTaylor} S.J. Parke and T.R. Taylor, Phys.\ Rev.\ Lett.\
         56:2459 (1986).
\bibitem {BerendsGiele} F.A. Berends and W.T. Giele,
         Nucl.\ Phys.\ B306:759 (1988).
\bibitem{SusyOne} Z. Bern, D. Dunbar, L. Dixon and D.A. Kosower,
         Nucl. Phys. B435:59 (1995).
\bibitem{AllPlus} Z. Bern, G. Chalmers, L. Dixon and D.A. Kosower,
         Phys. Rev. Lett. 72:2134 (1994).
\bibitem{BernChalmers} Z. Bern and G. Chalmers, hep-ph/9503236,
         to appear in Nucl.\ Phys. B.
\bibitem{PV} L.M.\ Brown and R.P.\ Feynman, Phys.\ Rev.\ 85:231
         (1952)\semi
         G.\ Passarino and M.\ Veltman, Nucl.\ Phys.\ B160:151
         (1979)\semi
         G. 't Hooft and M. Veltman, Nucl.\ Phys.\ B153:365 (1979).
\bibitem{Mahlon} G. Mahlon, Phys. Rev. D49:2197 (1994);
         Phys. Rev. D49:4438 (1994).
%
\end{thebibliography}
\end{document}